\documentclass[aps,floatfix,superscriptaddress,notitlepage]{
revtex4}
\usepackage{amsmath,amssymb,amsfonts,graphics,graphicx,dcolumn,bm,enumerate}
\usepackage{comment,natbib,appendix}
\usepackage{multirow,color}

\usepackage{amsthm}
\usepackage{epsfig}
\usepackage{natbib}
\usepackage{hyperref}
\usepackage{graphicx}
\usepackage{epstopdf}

\newcommand{\cmp}
{\affiliation{Condensed Matter Physics Division, 
Saha Institute of Nuclear Physics, 1/AF Bidhannagar, Kolkata 700064, India.}}
\newcommand{\isi}
{\affiliation{Economic Research Unit, Indian Statistical Institute, 203 B. T. Road, Kolkata 700108, India.}}
\newcommand{\aalto}
{\affiliation{Department of Computer Science, Aalto University School of 
Science, P.O. Box 15400, FI-00076 AALTO, Finland}}

\begin{document}

\title{Socio-economic inequality: Relationship between Gini and Kolkata indices}

\author{Arnab Chatterjee}%
\email[Email: ]{arnabchat@gmail.com} 
\cmp
\author{Asim Ghosh}
\email[Email: ]{asim.ghosh@aalto.fi}
\aalto
\author{Bikas K Chakrabarti}%
\email[Email: ]{bikask.chakrabarti@saha.ac.in}
\cmp \isi

\begin{abstract}
Socio-economic inequality is characterized from data using various indices.
The Gini ($g$) index, giving the overall inequality is the most common one, 
while the recently introduced Kolkata ($k$) index gives a measure of $1-k$ 
fraction of population who possess top  $k$ fraction of wealth in the society.
Here, we show the relationship between the two indices, using both empirical 
data and analytical estimates. The significance of their relationship has been 
discussed.
\end{abstract}

\maketitle

\section{Introduction}
Human social interactions often lead to complex dynamics.
Repeated social interactions produce spontaneous variations which are 
manifested as inequalities at various levels. 
The availability of huge amount of empirical data for a plethora of 
measures of human social interactions has made it possible to uncover 
the patterns, analyze them and look for the reasons behind various 
socio-economic inequalities.
Besides using tools of statistical physics, researchers are also bringing in 
knowledge and techniques from various other disciplines~\cite{lazer09}, e.g., 
statistics, applied mathematics, information theory and computer science 
to better the understanding of the precise nature (spatio-temporal) and  
origin of socio-economic inequalities prevalent in our society.

Socio-economic 
inequality~\cite{arrow2000meritocracy,stiglitz2012price,neckerman2004social,
goldthorpe2010analysing,chatterjee2015sociorev} 
basically concerns the existence of unequal `wealth' and `fortunes'
accumulated due to complex dynamics within the society. It usually 
contains structured and recurrent patterns of unequal distributions of goods, 
wealth, opportunities, and even rewards and punishments, and classically 
measured in terms of  \textit{inequality of conditions},
and \textit{inequality of opportunities}.
\textit{Inequality of conditions} refers to the unequal distribution of income,
wealth, assets and material goods. 
\textit{Inequality of opportunities} refers to the unequal distribution of `life
chances'. This is reflected in measures like level of education, health 
status, treatment by the criminal justice system etc. Socio-economic 
inequalities are responsible for conflict, war, crisis, oppression, criminal 
activities, political instability and unrest, and that indirectly affects 
economic growth~\cite{hurst1995social} of a region.
Traditionally, economic inequalities have been extensively studied in the 
context of income and 
wealth~\cite{yakovenko2009colloquium,chakrabarti2013econophysics,
aoyama2010econophysics},
although it is also measured for many quantities like energy 
consumption~\cite{lawrence2013global}.
The studies of inequality in 
society~\cite{piketty2014inequality,Cho23052014,Chin23052014,Xie23052014} has 
always been very important, and is also a topic of current focus and immediate 
global interest, bringing together researchers across various disciplines -- 
economics, sociology, mathematics, statistics, demography, geography, graph 
theory, computer science, and even theoretical physics.

Socio-economic inequalities are quantified in numerous ways. 
The most detailed measures are of course given by probability distributions of 
various quantities.
What is usually observed is that most quantities 
display broad distributions -- most common are log-normals, power-laws or their 
combinations.
For example, the distribution of income is usually an exponential followed by a power 
law~\cite{druagulescu2001exponential,chakrabarti2013econophysics}.
However, such distributions can widely differ in their forms and 
subtleties, and as such they are not quite convenient to handle. This lead to 
the introduction of various indices like the Gini~\cite{gini1921measurement}, 
Theil~\cite{theil1967economics}, Pietra~\cite{eliazar2010measuring} and other 
socio-geometric 
indices~\cite{eliazar2015asociogeometry,eliazar2015bsociogeometry}, which try 
to characterize various geometric features of these distributions.

\begin{figure}[t]
\centering
\includegraphics[width=7.0cm]{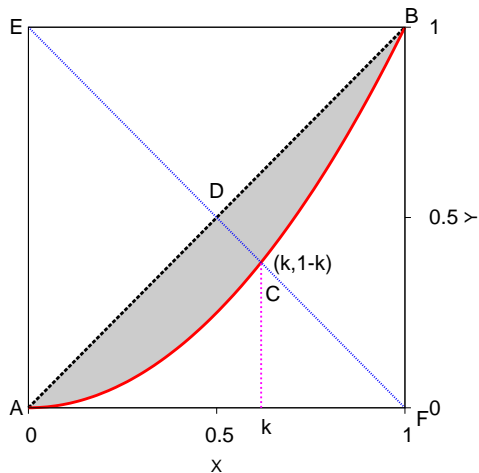}
 \caption{Lorenz curve (in solid red) for a typical probability distribution 
function and the equality line (dotted black diagonal).
Lorenz curve gives the cumulative fraction of `wealth' possessed by the 
corresponding fraction of poorer population.
 $g$-index is given by area of the shaded region (normalized by the area of the 
triangle AFB).
$k$-index is computed from the coordinate of the point of intersection C 
($k,1-k$) of the Lorenz curve and the diagonal perpendicular to 
the equality line.
Obviously, while $g$-index measures the overall inequality in the system, 
$k$-index gives the precise fraction $k$ of wealth possessed by $1-k$ fraction 
of richer population.
 }
 \label{fig:lorenz}
\end{figure}

The degree of inequality is most commonly measured by the Gini index. 
One considers the Lorenz curve~\cite{Lorenz}, representing
the cumulative proportion $X$ of ordered (from poorest to richest) individuals 
(entries) in terms of the cumulative  $Y$ of their wealth.
$Y$ can of course represent income or wealth of individuals but it can as well 
represent citation of articles, votes in favor of candidates, population of 
cities etc.
The Gini index ($g$), defined as the ratio of the area enclosed between 
the Lorenz curve and the equality line, to that below the equality line,
is the most common measure to quantify socio-economic inequality, taking values 
$0$ for absolute equality and $1$ for absolute inequality from a given 
statistical distribution.
If the area between 
(i) the Lorenz curve and the equality line is represented as $\cal{A}$, and 
(ii) that below the Lorenz curve as $\cal{B}$  (See Fig.~\ref{fig:lorenz}),
the Gini index is $g=\cal{A}/(\cal{A+B})=$ $2\cal{A}$.
Ghosh et al.~\cite{ghosh2014inequality}
recently introduced the Kolkata index (symbolizing 
the extreme nature of social inequalities in Kolkata) or `$k$-index', which is defined as the 
fraction $k$ such that poorest $(1-k)$ fraction of people possess $k$ fraction of 
income~\cite{inoue2015measuring,chatterjee2016universality,ghosh2016inequality}.
In fact, another recently proposed measure, the perpendicular width index 
$I_{PW}$~\cite{eliazar2015asociogeometry} can be shown to be equal to $\sqrt{2}(2k-1)$.

%

\section{Empirical findings on $g-k$ relationship}
\begin{figure*}[h]
\includegraphics[width=14.0cm]{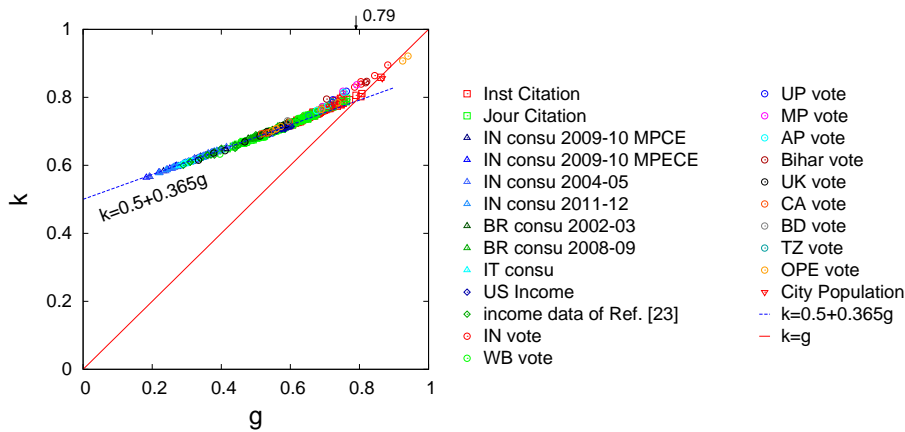}
\caption{Plot of the estimated values of $k$-index and $g$-index from various 
datasets (distributions): citations (retrieved from ISI Web of 
Science~\cite{ISI}, analyzed in 
Ref.~\cite{chatterjee2016universality}; Inst=institutions, Jour=journals)
expenditure (data from 
Ref.~\cite{chatterjee2016invariant,chakrabarti2016quantifying};
IN=India, BR=Brazil, IT=Italy), income (data 
from Ref.~\cite{inoue2015measuring}), voting data from proportional elections 
(data from Ref.~\cite{chatterjee2013universality}; OPE), 
voting data from first-past-the-post elections (data in the Appendix; 
IN=India, WB=West Bengal, UP=Uttar Pradesh, MP=Madhya Pradesh, AP=Andhra 
Pradesh, UK=United Kingdom, CA=Canada, BD=Bangladesh, TZ=Tanzania), and city 
population (data from Ref.~\cite{ghosh2014zipf}).
Data details are given in the Appendix.
The dotted straight line represents $k=0.5+0.365g$.
}
\label{fig:gk_emp}
\end{figure*}
A large variety of socio-economic data suggest that there exists 
a simple relation between the two seemingly different inequality measures. 
We analyzed citations of papers published from academic institutions and 
journals (data from ISI Web of Science~\cite{ISI} and reported in  
Ref.~\cite{chatterjee2016universality}), consumption expenditure data of 
India~\cite{NSSO}, Brazil~\cite{Brazil0203,Brazil0809}, Italy~\cite{Italydata},
income data from USA~\cite{IRS}, voting data from open list proportional 
elections~\cite{chatterjee2013universality} of Italy, Netherlands and Sweden,
\textit{first past the post} election data for Indian Parliamentary elections 
and Legislative Assembly elections~\cite{ECI}, 
United Kingdom~\cite{UKElec}, Canada~\cite{CAElec}, Bangladesh~\cite{BDElec}, 
Tanzania~\cite{TZElec}, and city population data from Ref.~\cite{ghosh2014zipf}.
See Tables I-IX in the Appendix for details.

The relation is perfectly linear for smaller values while the curve becomes 
non-linear as $g$ or $k$ approaches unity, the limit of extreme inequality 
(Fig.~\ref{fig:gk_emp}).
The most intriguing part is that the data from a variety of sources hardly 
depart from this smooth curve. We explore a spectrum of data such as income, 
expenditure, journal citations and impact factors, votes, city population to 
arrive at this 
conclusion.

The $k$-index and $g$-index show a linear relationship
\begin{equation}
 k= \frac{1}{2} + \gamma.g, \; \textrm{for} \; 0 \le g \lesssim 0.70,
 \label{eq:emp}
\end{equation}
with $\gamma =0.365 \pm 0.005$.

\section{Approximate analytical estimates}
\label{sec:analytical}
In Fig.~\ref{fig:lorenz}, the thick red line is a typical Lorenz curve 
corresponding to a probability distribution function $y=P(x)$.
$X$ denotes the cumulative share of $x$ from lowest to highest $y$ while $Y$ 
denotes the cumulative share of $y$.
The Lorenz curve cuts the anti-diagonal $Y=1-X$  at point C $(k,1-k)$ and thus 
the $k$-index is defined as the following (in terms of wealth posessed by 
individuals, say): $k$ fraction of total wealth is possessed by $1-k$ fraction of the top 
wealthiest people.
The Gini index $g$ is simply given by $2\mathcal{A}$.
If $\mathcal{A}$ is the shaded area enclosed by the Lorenz curve (ACB) and the 
equality line ADB ($Y=X$), then Gini coefficient $g$ is given by
\begin{equation}
 g= 
 \frac{\textrm{area of the shaded region}}{\textrm{area of the traingle ABE}}
 = 2 \mathcal{A}.
\end{equation}
We discusss below three approximate ways to calculate $\cal{A}$.

\subsubsection*{Case I: Lorenz curve as the broken straight lines AC \& CB}
From Fig.~\ref{fig:lorenz}, 
AB=$\sqrt{2}$ and CD=$\frac{\sqrt{2}}{2} - 
\sqrt{2}(1-k)=\frac{1}{\sqrt{2}} (2k-1)$. Thus the area of the triangle CAB is 
$\mathcal{A}_1=\frac{1}{2}$AB.CD $=\frac{1}{2}.\sqrt{2}. \frac{1}{\sqrt{2}} 
(2k-1) = \frac{1}{2} (2k-1)$.
Thus, 
\begin{equation}
g \ge 2\mathcal{A}_1 = 2k-1,
\label{eq:a1}
\end{equation}
giving
\begin{equation}
k \le \frac{1}{2} + \frac{1}{2} g.
\label{eq:kg_a1}
\end{equation}
It may be noted that the equality in the above relation corresponds to 
$g=k$ for $g=k=1$.

\subsubsection*{Case II: Lorenz curve as a straight line parallel to ADB at 
perpendicular distance DC}

Here, area $\mathcal{A}_2 =$ AB.CD $=\sqrt{2}.\frac{1}{\sqrt{2}}(2k-1)=2k-1$. 
This gives $g \le 2(2k-1)$ or 
\begin{equation}
 k \ge \frac{1}{2} + \frac{1}{4}g.
 \label{eq:kg_a2}
\end{equation}
In this approximation, the equality in the above relation corresponds to $g=k$ 
for $g=k=\frac{2}{3}$. Analysis of the observed data suggests that $k-g$ line 
(Eq.~\ref{eq:emp}; Fig.~\ref{fig:gk_emp}) touches $k=g$ line at around $0.78$.

\subsubsection*{Case III: Lorenz curve as an arc of a circle}
\begin{figure}[t]
\centering
\includegraphics[height=4.0cm]{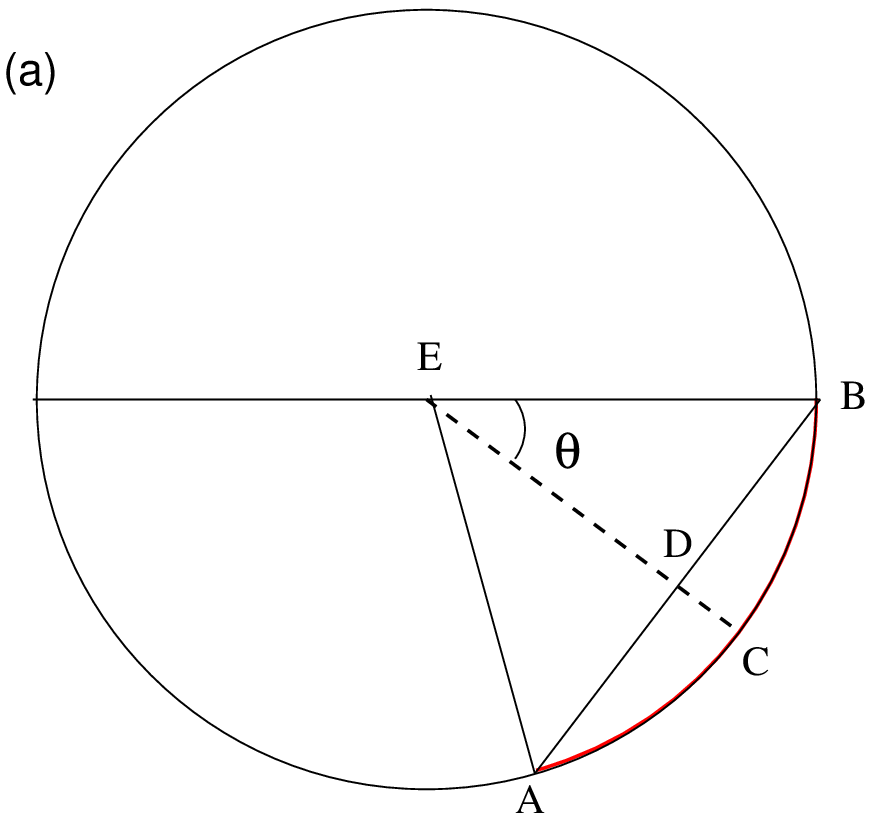}
\includegraphics[height=4.6cm]{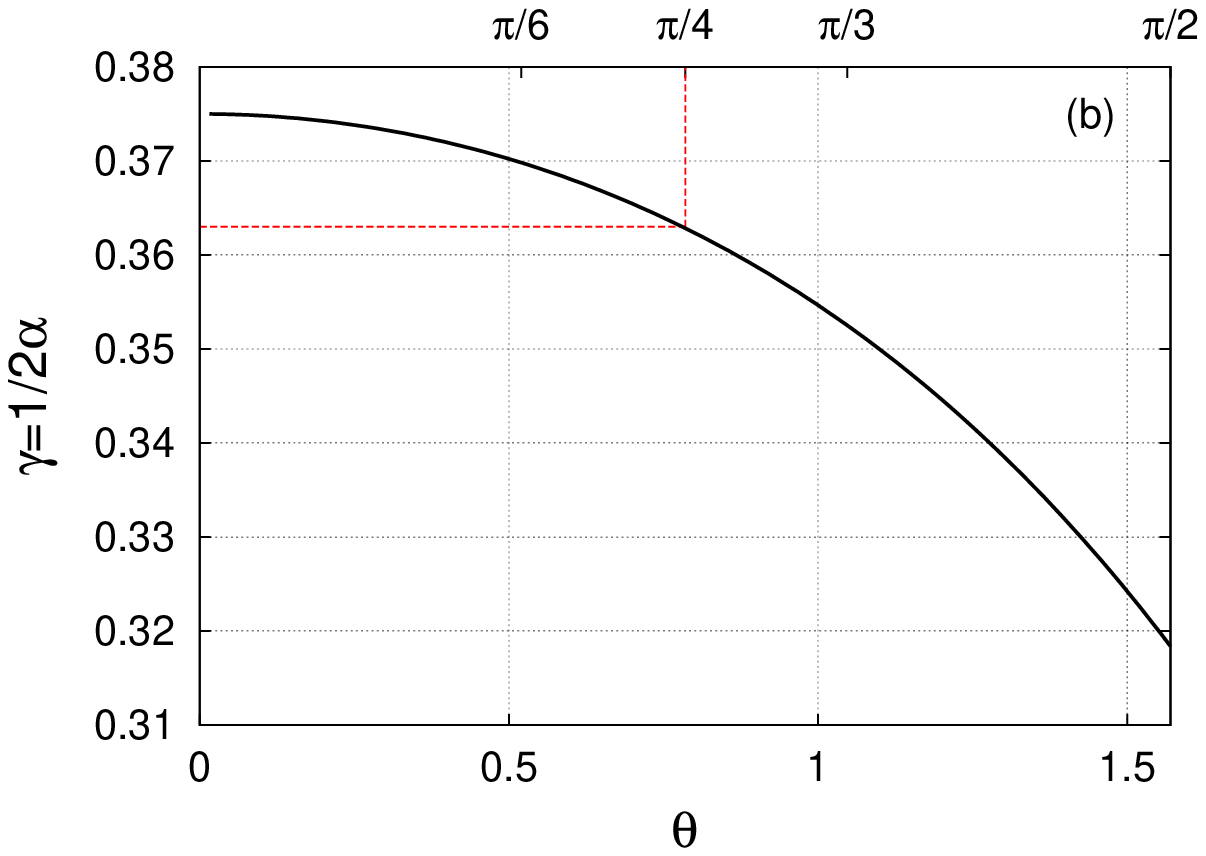}
 \caption{(a) Approximating the Lorenz curve (red line) as an arc of a 
circle and the equality line as the chord AB. The extremities of the arc ACB 
subtend an angle $2\theta$ at the centre of the circle E.
(b) Plot of $\gamma =\frac{1}{2\alpha}$ from 
Eq.~(\ref{eq:c}) for different values of $\theta$.
 }
 \label{fig:cvstheta}
\end{figure}

Let us now imagine that the Lorenz curve is represented by the arc ACB 
of a circle (Fig.~\ref{fig:cvstheta}a) of radius $r$ (=AE=BE).
DE is perpendicular to AB such that 
$\angle{BED} = \theta$. The total area of the sector BEAC is then $\theta r^2$.
The area of the triangle ABE is given by 
$\frac{1}{2}$.DE.AB = $\frac{1}{2}. r \cos \theta . 2r \sin \theta 
=r^2 \cos \theta \sin \theta $.
Thus our required area ACDB is given by (difference between the sector and the triangle 
defined above)
\begin{equation}
\mathcal{A}^\prime= r^2 (\theta - \sin \theta \cos \theta).
\end{equation}
If we write $\mathcal{A}=\mathcal{A}^\prime = \frac{\alpha}{2}$.AB.CD, then
\begin{equation}
 \alpha =\frac{\theta - 
\sin \theta \cos \theta}{\sin \theta (1- \cos \theta)}.
\label{eq:c}
\end{equation}
Referring back to Case I, and incorporating the 
factor $\alpha$, we get the approximate value $\cal{A}^\prime$ 
as $\alpha \mathcal{A}_1$. 
Hence $g= 2 \alpha \mathcal{A}_1 = \alpha (2k-1)$ (using Eq.~\ref{eq:a1}). This 
gives
\begin{equation}
 k=\frac{1}{2} + \frac{1}{2\alpha}g.
 \label{eq:kg_c}
\end{equation}
Thus the slope of the $k-g$ line is $\gamma=\frac{1}{2\alpha}$.
Variation of $\frac{1}{2\alpha}$ with $\theta$  is plotted in 
Fig.~\ref{fig:cvstheta}b.
The observed approximate value of $\gamma$ (From Fig.~\ref{fig:gk_emp}) is 
$0.363$ which corresponds to $\theta = \pi/4$ (see Fig.~\ref{fig:cvstheta}b).
This would imply that the Lorenz curve can be approximated as a quadrant arc 
of a circle with centre at E (see Fig.~\ref{fig:cvstheta}a), subtending and 
angle $2\theta = \pi/2$ at E (compare Fig.~\ref{fig:cvstheta}a with 
Fig.~\ref{fig:lorenz}).
In that case, the $g-k$ line will touch the $k=g$ line at around $0.78$ (from 
Eq.~\ref{eq:kg_c}).

In fact, this linear relationship (with the value of the slope $\gamma \approx 
0.363$) derived here for a circular (quadrant) Lorenz curve is more generally 
valid. If the Lorenz curve $L(x)$ in Fig.~\ref{fig:lorenz} is taken as a 
parabola ($L(x)=x^2$ for uniform  and normalized distribution $P(m)$ of
income/wealth $m$; 
$L(x)=\int_0^x 2m P(m) dm$), one gets $g=2\int_0^1 (x-L(x))dx =\frac{1}{3} 
\approx 0.33$ and $1-k=L(k)=k^2$, giving $k=\frac{1}{2}(\sqrt{5} -1) \approx 
0.62$. These values of $g$ and $k$ satisfy the above relationship (Eq.~\ref{eq:emp}) very well.

\begin{figure}[!h]
\centering
\includegraphics[width=5.5cm]{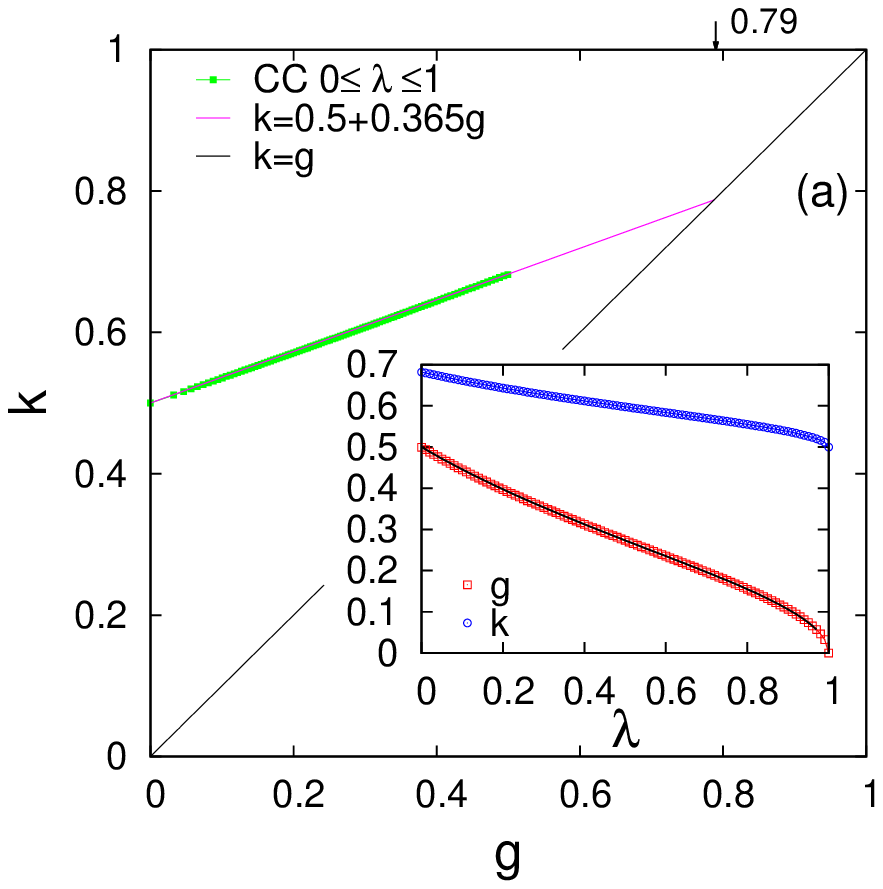}
\includegraphics[width=5.5cm]{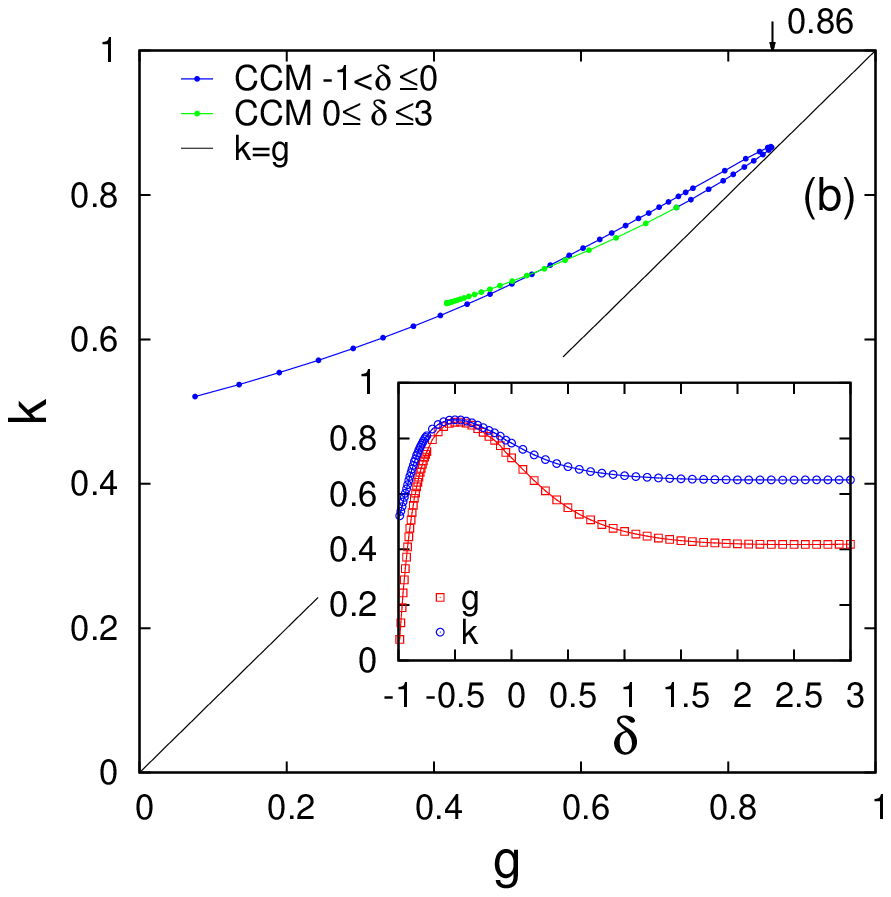}
 \caption{Monte Carlo simulation results for $g$ vs. $k$ in CC and CCM models 
(for $1000$ agents). 
 (a) For CC model, varying parameter $\lambda$.
 The inset shows the plots of $g$ and $k$ in the range of $0\le \lambda \le 1$. 
 The solid line corresponds to Eq.~\ref{eq:gpck} for $g$ vs. $\lambda$ which agrees perfectly 
with the results from simulations.
 In the $g-k$ plot, the points fit to $k= \frac{1}{2} + \gamma.g$ with $\gamma \approx 0.365 
\pm 0.005$.
 (b)  For CCM model, varying parameter $\delta$. The inset shows the 
variation of $g$ and $k$ in the range of $-1 < \delta \le 3$.
 }
 \label{fig:gk_models}
\end{figure}

\section{Estimates of $g-k$ relation from kinetic exchange models}
\label{sec:model}
Let us now consider some market models developed by econophysicists, in 
particular the kinetic exchange 
models~\cite{Chatterjee2007,chakrabarti2013econophysics}. In the CC 
model~\cite{Chatterjee2007} there, an agent keeps a fraction $\lambda$ (same 
for all) of their income or wealth before going for any (stochastic) exchange 
(trade or scattering) with another agent. 
Formally, the dynamics is defined by
\begin{equation*}
\begin{aligned}
m_i(t+1) &= \lambda m_i(t)+r(1-\lambda)\left[ m_i(t)+m_j(t) \right] \\
m_j(t+1) &= \lambda m_j(t)+(1-r)(1-\lambda)\left[ m_i(t)+m_j(t) \right],
\end{aligned}
\label{eq:cc}
\end{equation*}
where $r$ is a random fraction in $[0,1]$, drawn in each time step 
(trade or exchange). 
$m_i(t)$ and $m_i(t+1)$ are the wealth of the $i$th agent at trading times $t$ 
and $(t+1)$ respectively.
The steady state distribution of 
wealth is Gamma like~\cite{Patriarca2004,Chatterjee2007} with the peak position 
shifting to higer income or wealth with increasing $\lambda$ (Gibbs or 
exponential distribution for $\lambda=0$ and $\delta$-function for $\lambda 
\to 1$). 
In fact, the distributions fit 
to~\cite{Patriarca2004,chakrabarti2010inequality,chakrabarti2013econophysics}
\begin{equation}
f_n(m) = \frac{1}{\Gamma(n)} \left( \frac{n}{\langle m \rangle} \right)^n 
m^{n-1} \exp\left( -\frac{nm}{\langle m \rangle} \right),
\label{eq:pck1}
\end{equation}
with 
\begin{equation}
n(\lambda) = 1+ \frac{3\lambda}{1-\lambda}. 
\label{eq:pck2}
\end{equation}
Eq.~\ref{eq:pck1} is a standard Gamma distribution whose Gini index is given by
\begin{equation}
 g = \frac{\Gamma \left( n + \frac{1}{2} \right)}{\sqrt{\pi} n \Gamma(n) },
 \label{eq:gpck}
\end{equation}
with $n$ given by Eq.~\ref{eq:pck2}.

$g$ and $k$ computed for wealth distributions of CC model~\cite{ghosh2016inequality} using 
numerical simulations are given in inset of Fig.~\ref{fig:gk_models}a. For Gini index, we also 
plotted Eq.~\ref{eq:gpck} and found to coincide with the  results from numerical simulation.
The $g-k$ relationship is also 
found to be linear~Fig.~\ref{fig:gk_models}a, obeying $k= \frac{1}{2} + 
\gamma.g$ with $\gamma \approx 0.365 \pm 0.005$.
This compares very well with the $g-k$ relationship derived in 
Sec.~\ref{sec:analytical} Case III.

In the CCM model~\cite{Chatterjee2007,chakrabarti2013econophysics}, each agent 
$i$ has a saving fraction $\lambda$ drawn from a 
(quenched) distribution $\Pi(\lambda) = (1+\delta)(1-\lambda)^\delta$. 
Following similar stochastic dynamics as in CC model, 
\begin{equation*}
\begin{aligned}
m_i(t+1) &= \lambda_i m_i(t)+r \left[ (1-\lambda_i)m_i(t)+(1-\lambda_j)m_j(t) 
\right] \\
m_j(t+1) &= \lambda_j 
m_j(t)+(1-r)\left[ (1-\lambda_i)(m_i(t)+(1-\lambda_j)m_j(t) \right],
\end{aligned}
\label{eq:ccm}
\end{equation*}
one gets a steady state distribution of income or wealth with power law tails 
$P(m)\sim m^{-(2+\delta)}$ for large $m$~\cite{Chatterjee2007}.
$g$ and $k$ computed for such distributions~\cite{ghosh2016inequality} are 
given in inset of Fig.~\ref{fig:gk_models}b for varying range of $\delta$. The 
$g-k$ relationship here is found to be nonlinear (see 
Fig.~\ref{fig:gk_models}b) but 
very much  around a similar linear relationship.

\section{Discussions}
As already emphasized, the Gini index $g$ is the most popular among economists 
and sociologists. It gives an overall measure of the inequality in a 
society. As can be seen from Fig~\ref{fig:lorenz}, it requires accurate data 
for the entire Lorenz curve to give a measure of the shaded area enclosed by it 
and the equality line. However, the data for the low income group as well as 
the high income group in the society are not always very easy to obtain. 
The Lorenz curve being determined by the cumulative distribution, estimates of 
both $g$ and $k$ indices are affected. Of course, the Kolkata index $k$ 
being given by the intersection point of the Lorenz curve and the diagonal 
perpendicular to the equality line, where the data are usually expected to be 
rather accurate and massive, the $k$-index value should be less affected compared to the 
$g$-index which is rather directly affected by the lack of proper data.
Indeed as shown in Sec.~\ref{sec:analytical}, the $g-k$ linear relationship is 
extremely robust and fits different forms of Lorenz curve and therefore,  
distributions of income, wealth, citations, etc. This robustness is also 
observed empirically (Fig.~\ref{fig:gk_emp}).
Hence the $g-k$ relationship studied here would be 
extremely useful to translate from one inequality measure to the other; since 
$1-k$ fraction of people possess precisely $k$ fraction of the total 
wealth, translation of social inequality measures into $k$-index language can  
be of major significance.

\section*{acknowledgments}
 We are grateful to Parthasarathi Mitra for his help with 
Sec.~\ref{sec:analytical}, and some useful comments.
We also thank our referees for very useful suggestions (added in Sec.~\ref{sec:model} 
and Appendix A).

%
%

\newpage

\appendix
\section{Appendix: Using $2k-1$ instead of $k$}
\begin{figure*}[h]
\includegraphics[width=14.0cm]{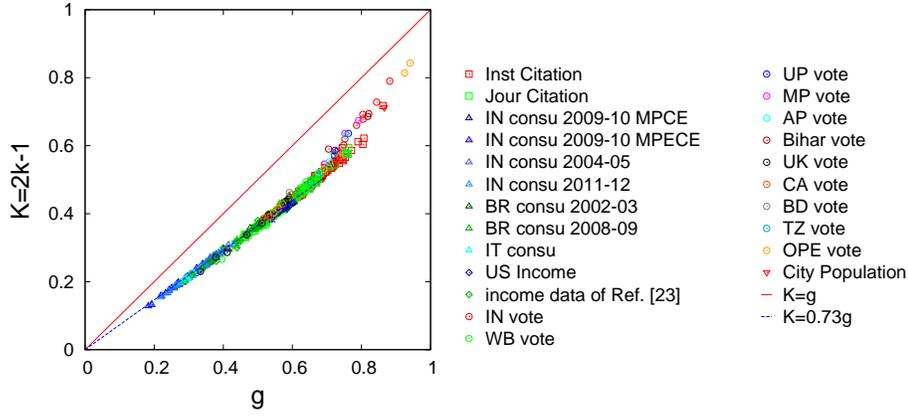}
\caption{Plot of the estimated values of $k$-index as $K=2k-1$ and $g$-index from various 
datasets (distributions) as plotted in Fig.~\ref{fig:gk_emp}. The solid red straight like is 
$K=g$, while dotted blue straight line represents $K=0.73g$.
}
\label{fig:gk_emp1}
\end{figure*}

An alternative way of plotting the $k$ index is to consider the quantity $K=2k-1$, which will 
be defined now in $[0,1]$ by definition. Fig~\ref{fig:gk_emp1} shows this plot with the 
strict inequality line $2k-1=K=g$, which is never exceeded since $g \ge 2k-1$ 
(Eq.~\ref{eq:a1}). The transformation to $K$ makes the slope of the $K$ vs $g$ plot for 
smaller $g$ values equal to $2\gamma \approx 0.73$.
\newpage 

\section{Appendix: Computed $g$ and $k$ indices from data for citations, 
income, expenditure, vote and city size.}
\begin{table}[!htbp] \tiny
\caption{Computed $g$ and $k$ indices for academic institutions from 
citation data for different years. 
Data retreived from ISI Web of Science~\cite{ISI} and available from 
Ref.~\cite{chatterjee2016universality}.}
\label{tab:inst}
\begin{tabular}{|c|c|c|c|}
\hline

Institutions  & Year  & $g$ & $k$ \\ 
\hline

Bern & 1980 & 0.705 & 0.767 \\
& 1990 & 0.684 & 0.759 \\
& 2000 & 0.615 & 0.726 \\
& 2010 & 0.621 & 0.726 \\ \hline
BHU & 1980 & 0.681 & 0.757 \\
& 1990 & 0.708 & 0.767 \\
& 2000 & 0.635 & 0.737 \\
& 2010 & 0.628 & 0.728 \\ \hline
Bordeaux & 1980 & 0.685 & 0.762 \\
& 1990 & 0.671 & 0.750 \\
& 2000 & 0.647 & 0.740 \\
& 2010 & 0.606 & 0.723 \\ \hline
Boston & 1980 & 0.661 & 0.749 \\
& 1990 & 0.690 & 0.759 \\
& 2000 & 0.657 & 0.744 \\
& 2010 & 0.620 & 0.729 \\ \hline
Bristol & 1980 & 0.599 & 0.717 \\
& 1990 & 0.642 & 0.737 \\
& 2000 & 0.637 & 0.736 \\
& 2010 & 0.607 & 0.722 \\ \hline
Buenos Aires & 1980 & 0.804 & 0.802 \\
& 1990 & 0.634 & 0.734 \\
& 2000 & 0.654 & 0.741 \\
& 2010 & 0.669 & 0.746 \\ \hline
Calcutta & 1980 & 0.697 & 0.768 \\
& 1990 & 0.626 & 0.735 \\
& 2000 & 0.650 & 0.738 \\
& 2010 & 0.571 & 0.710 \\ \hline
Caltech & 1980 & 0.671 & 0.751 \\
& 1990 & 0.654 & 0.743 \\
& 2000 & 0.650 & 0.742 \\
& 2010 & 0.642 & 0.737 \\ \hline
Cambridge & 1980 & 0.697 & 0.762 \\
& 1990 & 0.715 & 0.770 \\
& 2000 & 0.674 & 0.752 \\
& 2010 & 0.651 & 0.741 \\ \hline
Chicago & 1980 & 0.646 & 0.742 \\
& 1990 & 0.656 & 0.746 \\
& 2000 & 0.664 & 0.747 \\
& 2010 & 0.666 & 0.747 \\ \hline
Cologne & 1980 & 0.680 & 0.755 \\
& 1990 & 0.712 & 0.768 \\
& 2000 & 0.644 & 0.740 \\
& 2010 & 0.607 & 0.723 \\ \hline
Columbia & 1980 & 0.661 & 0.747 \\
& 1990 & 0.655 & 0.743 \\
& 2000 & 0.644 & 0.741 \\
& 2010 & 0.632 & 0.733 \\ \hline
Delhi & 1980 & 0.645 & 0.741 \\
& 1990 & 0.675 & 0.757 \\
& 2000 & 0.667 & 0.749 \\
& 2010 & 0.615 & 0.724 \\ \hline
Edinburgh & 1980 & 0.721 & 0.772 \\
& 1990 & 0.654 & 0.742 \\
& 2000 & 0.652 & 0.742 \\
& 2010 & 0.648 & 0.740 \\ \hline
Gottingen & 1980 & 0.644 & 0.739 \\
& 1990 & 0.807 & 0.811 \\
& 2000 & 0.657 & 0.744 \\
& 2010 & 0.633 & 0.734 \\ \hline
Groningen & 1980 & 0.620 & 0.730 \\
& 1990 & 0.642 & 0.737 \\
& 2000 & 0.612 & 0.725 \\
& 2010 & 0.590 & 0.717 \\ \hline
Harvard & 1980 & 0.695 & 0.761 \\
& 1990 & 0.712 & 0.769 \\
& 2000 & 0.667 & 0.750 \\
& 2010 & 0.641 & 0.738 \\ \hline
Heidelberg & 1980 & 0.638 & 0.738 \\
& 1990 & 0.640 & 0.739 \\
& 2000 & 0.645 & 0.741 \\
& 2010 & 0.647 & 0.742 \\ \hline
Helsinki & 1980 & 0.634 & 0.736 \\
& 1990 & 0.627 & 0.733 \\
& 2000 & 0.626 & 0.733 \\
& 2010 & 0.626 & 0.733 \\ \hline
HUJ & 1980 &  0.633 & 0.732 \\
& 1990 & 0.639 & 0.739 \\
& 2000 & 0.656 & 0.743 \\
& 2010 & 0.594 & 0.717 \\ \hline
IISC & 1980 & 0.715 & 0.771 \\
& 1990 & 0.699 & 0.764 \\
& 2000 & 0.657 & 0.746 \\
& 2010 & 0.606 & 0.722 \\ \hline

%

\end{tabular}
\begin{tabular}{|c|c|c|c|}
\hline

Institutions  & Year  & $g$ & $k$ \\ 
\hline

Kyoto & 1980 & 0.651 & 0.743 \\
& 1990 & 0.662 & 0.747 \\
& 2000 & 0.668 & 0.749 \\
& 2010 & 0.619 & 0.728 \\ \hline
Landau & 1980 & 0.862 & 0.859 \\
& 1990 & 0.745 & 0.797 \\
& 2000 & 0.790 & 0.806 \\
& 2010 & 0.619 & 0.733 \\ \hline
Leiden & 1980 & 0.612 & 0.726 \\
& 1990 & 0.617 & 0.727 \\
& 2000 & 0.616 & 0.727 \\
& 2010 & 0.617 & 0.727 \\ \hline
Leuven & 1980 & 0.662 & 0.747 \\
& 1990 & 0.692 & 0.761 \\
& 2000 & 0.654 & 0.744 \\
& 2010 & 0.638 & 0.737 \\ \hline
Madras & 1980 & 0.666 & 0.754 \\
& 1990 & 0.666 & 0.756 \\
& 2000 & 0.622 & 0.728 \\
& 2010 & 0.753 & 0.783 \\ \hline
Manchester & 1980 & 0.666 & 0.756 \\
& 1990 & 0.753 & 0.783 \\
& 2000 & 0.665 & 0.747 \\
& 2010 & 0.622 & 0.728 \\ \hline
Melbourne & 1980 & 0.572 & 0.710 \\
& 1990 & 0.595 & 0.717 \\
& 2000 & 0.606 & 0.719 \\
& 2010 & 0.622 & 0.728 \\ \hline
MIT & 1980 & 0.713 & 0.769 \\
& 1990 & 0.724 & 0.777 \\
& 2000 & 0.716 & 0.772 \\
& 2010 & 0.687 & 0.759 \\ \hline
Osaka & 1980 & 0.624 & 0.732 \\
& 1990 & 0.703 & 0.764 \\
& 2000 & 0.646 & 0.742 \\
& 2010 & 0.680 & 0.753 \\ \hline
Oslo & 1980 & 0.658 & 0.744 \\
& 1990 & 0.647 & 0.740 \\
& 2000 & 0.603 & 0.721 \\
& 2010 & 0.587 & 0.715 \\ \hline
Oxford & 1980 & 0.647 & 0.742 \\
& 1990 & 0.692 & 0.761 \\
& 2000 & 0.687 & 0.756 \\
& 2010 & 0.665 & 0.747 \\ \hline
Princeton & 1980 & 0.757 & 0.788 \\
& 1990 & 0.743 & 0.784 \\
& 2000 & 0.714 & 0.767 \\
& 2010 & 0.683 & 0.753 \\ \hline
SINP & 1980 & 0.670 & 0.746 \\
& 1990 & 0.632 & 0.731 \\
& 2000 & 0.648 & 0.741 \\
& 2010 & 0.679 & 0.752 \\ \hline
Stanford & 1980 & 0.737 & 0.780 \\
& 1990 & 0.698 & 0.764 \\
& 2000 & 0.734 & 0.779 \\
& 2010 & 0.679 & 0.755 \\ \hline
Stockholm & 1980 & 0.695 & 0.762 \\
& 1990 & 0.664 & 0.752 \\
& 2000 & 0.685 & 0.756 \\
& 2010 & 0.698 & 0.762 \\ \hline
TAU & 1980 & 0.718 & 0.770 \\
& 1990 & 0.679 & 0.751 \\
& 2000 & 0.663 & 0.746 \\
& 2010 & 0.657 & 0.744 \\ \hline
TIFR & 1980 & 0.699 & 0.765 \\
& 1990 & 0.745 & 0.780 \\
& 2000 & 0.736 & 0.774 \\
& 2010 & 0.747 & 0.778 \\ \hline
Tokyo & 1980 & 0.666 & 0.748 \\
& 1990 & 0.677 & 0.754 \\
& 2000 & 0.676 & 0.752 \\
& 2010 & 0.655 & 0.743 \\ \hline
Toronto & 1980 & 0.771 & 0.793 \\
& 1990 & 0.714 & 0.769 \\
& 2000 & 0.684 & 0.756 \\
& 2010 & 0.649 & 0.739 \\ \hline
Yale & 1980 & 0.716 & 0.770 \\
& 1990 & 0.725 & 0.774 \\
& 2000 & 0.723 & 0.773 \\
& 2010 & 0.684 & 0.756 \\  \hline
Zurich & 1980 & 0.718 & 0.773 \\
& 1990 & 0.684 & 0.756 \\
& 2000 & 0.661 & 0.748 \\
& 2010 & 0.629 & 0.729 \\  \hline

\end{tabular}
\end{table}

\begin{table}[!htbp]\scriptsize
\caption{Computed $g$ and $k$ indices for academic journals for citation data 
for different years.
Data retreived from ISI Web of Science~\cite{ISI} and available from 
Ref.~\cite{chatterjee2016universality}.}
\label{tab:inst}
\begin{tabular}{|c|c|c|c|}
\hline

Institutions  & Year  & $g$ & $k$ \\ 
\hline

Astronomy & 1980 & 0.636 & 0.734 \\
Astrophys. & 1990 & 0.577 & 0.715 \\
& 2000 & 0.558 & 0.704 \\
& 2010 & 0.564 & 0.704 \\ \hline
Astrophys. & 1980 & 0.550 & 0.701 \\
J.& 1990 & 0.533 & 0.696 \\
& 2000 & 0.547 & 0.701 \\
& 2010 & 0.506 & 0.685 \\ \hline
Biochem. & 1980 & 0.486 & 0.674 \\
& 1990 & 0.509 & 0.684 \\
& 2000 & 0.435 & 0.656 \\
& 2010 & 0.436 & 0.654 \\ \hline
BMJ & 1980 & 0.676 & 0.757 \\
& 1990 & 0.692 & 0.763 \\
& 2000 & 0.709 & 0.769 \\
& 2010 & 0.507 & 0.682 \\ \hline
Circulation & 1980 & 0.555 & 0.704 \\
& 1990 & 0.571 & 0.713 \\
& 2000 & 0.528 & 0.693 \\
& 2010 & 0.492 & 0.675 \\ \hline
CPL & 1980 & 0.606 & 0.719 \\
& 1990 & 0.627 & 0.730 \\
& 2000 & 0.579 & 0.713 \\
& 2010 & 0.525 & 0.687 \\ \hline
Eur. J. & 1980 & 0.545 & 0.697 \\ 
Biochem. & 1990 & 0.531 & 0.693 \\
& 2000 & 0.514 & 0.683 \\
& 2010 & 0.545 & 0.698 \\ \hline
Inor. Chem. & 1980 & 0.459 & 0.666 \\
& 1990 & 0.476 & 0.672 \\
& 2000 & 0.466 & 0.668 \\
& 2010 & 0.447 & 0.662 \\ \hline
JAMA & 1980 & 0.675 & 0.753 \\
& 1990 & 0.762 & 0.787 \\
& 2000 & 0.757 & 0.788 \\
& 2010 & 0.723 & 0.772 \\ \hline
JAP & 1980 & 0.668 & 0.754 \\
& 1990 & 0.638 & 0.739 \\
& 2000 & 0.613 & 0.727 \\
& 2010 & 0.511 & 0.685 \\ \hline
J. Chem.& 1980 & 0.651 & 0.739 \\
Phys. & 1990 & 0.582 & 0.711 \\
& 2000 & 0.579 & 0.710 \\
& 2010 & 0.522 & 0.686 \\ \hline
JMMM & 1980 & 0.631 & 0.735 \\
& 1990 & 0.653 & 0.744 \\
& 2000 & 0.584 & 0.714 \\
& 2010 & 0.570 & 0.708 \\ \hline
J. Org.& 1980 & 0.513 & 0.687 \\
Chem. & 1990 & 0.494 & 0.680 \\
& 2000 & 0.442 & 0.659 \\
& 2010 & 0.417 & 0.649 \\ \hline
JPA & 1980 & 0.752 & 0.790 \\
& 1990 & 0.625 & 0.735 \\
& 2000 & 0.592 & 0.722 \\
& 2010 & 0.573 & 0.707 \\ \hline
Lancet & 1980 & 0.650 & 0.736 \\
& 1990 & 0.604 & 0.721 \\
& 2000 & 0.642 & 0.739 \\
& 2010 & 0.463 & 0.670 \\ \hline
%
%

\end{tabular}
\begin{tabular}{|c|c|c|c|}
\hline

Institutions  & Year  & $g$ & $k$ \\ 
\hline

Macromol. & 1980 & 0.642 & 0.737 \\
& 1990 & 0.567 & 0.710 \\
& 2000 & 0.499 & 0.682 \\
& 2010 & 0.472 & 0.668 \\ \hline
Nature & 1980 & 0.637 & 0.736 \\
& 1990 & 0.676 & 0.751 \\
& 2000 & 0.668 & 0.746 \\
& 2010 & 0.547 & 0.698 \\ \hline
NEJM & 1980 & 0.518 & 0.686 \\
& 1990 & 0.506 & 0.683 \\
& 2000 & 0.603 & 0.720 \\
& 2010 & 0.576 & 0.706 \\ \hline
Physica A & 1980 & 0.551 & 0.700 \\
& 1990 & 0.653 & 0.748 \\
& 2000 & 0.649 & 0.744 \\
& 2010 & 0.587 & 0.718 \\ \hline
Tetrahedron & 1980 & 0.709 & 0.771 \\
& 1990 & 0.556 & 0.701 \\
& 2000 & 0.503 & 0.680 \\
& 2010 & 0.465 & 0.665 \\ \hline
Physica B & 1990 & 0.632 & 0.732 \\
& 2000 & 0.647 & 0.740 \\
& 2010 & 0.558 & 0.702 \\ \hline
Physica C & 1990 & 0.586 & 0.715 \\
& 2000 & 0.664 & 0.748 \\
& 2010 & 0.658 & 0.744 \\ \hline
PRA & 1980 & 0.609 & 0.724 \\
& 1990 & 0.603 & 0.724 \\
& 2000 & 0.624 & 0.729 \\
& 2010 & 0.519 & 0.687 \\ \hline
PRB & 1980 & 0.648 & 0.743 \\
& 1990 & 0.649 & 0.741 \\
& 2000 & 0.602 & 0.722 \\
& 2010 & 0.528 & 0.692 \\ \hline
PRC & 1980 & 0.653 & 0.744 \\
& 1990 & 0.617 & 0.728 \\
& 2000 & 0.569 & 0.709 \\
& 2010 & 0.545 & 0.697 \\ \hline
PRD & 1980 & 0.763 & 0.797 \\
& 1990 & 0.681 & 0.759 \\
& 2000 & 0.613 & 0.728 \\
& 2010 & 0.532 & 0.693 \\ \hline
PRE & 2000 & 0.584 & 0.715 \\
& 2010 & 0.492 & 0.678 \\ \hline
PRL & 1980 & 0.670 & 0.746 \\
& 1990 & 0.604 & 0.724 \\
& 2000 & 0.589 & 0.717 \\
& 2010 & 0.493 & 0.679 \\ \hline
Science & 1980 & 0.635 & 0.738 \\
& 1990 & 0.663 & 0.745 \\
& 2000 & 0.614 & 0.725 \\
& 2010 & 0.529 & 0.692 \\ \hline
Langmuir & 1990 & 0.589 & 0.716 \\
& 2000 & 0.528 & 0.694 \\
& 2010 & 0.460 & 0.665 \\ \hline
&  &  &  \\
&  &  &  \\
&  &  &  \\
&  &  &  \\
&  &  &  \\
\hline
\end{tabular}
\end{table}

\begin{table}
\caption{Estimated values of $g$ and $k$-indices for per capita consumption 
expenditure for India for MPCE \& MPECE, 2009-2010:  Data is taken 
from Ref.~\cite{NSSO} and Ref.~\cite{chatterjee2016invariant}.}
\label{tab:in_consu_g_k_1}

\begin{tabular}{|c|c|c|c|c|c|}
\hline

State &  \multicolumn{2}{c|}{MPCE}   & 
\multicolumn{2}{c|}{MPECE}  \\ \cline{2-5} 
& $g$ & $k$ &  $g$ & $k$  
\\ \hline 

Jammu \& Kashmir & 0.277 & 0.598 & 0.258 & 0.591 \\  \hline 
Himachal Pradesh & 0.356 & 0.628 & 0.308 & 0.609 \\ \hline 
Punjab & 0.342 & 0.623 & 0.321 & 0.616 \\ \hline 
Chandigar & 0.401 & 0.648 & 0.378 & 0.639 \\ \hline 
Uttaranchal & 0.324 & 0.615 & 0.273 & 0.596 \\ \hline 
Haryana & 0.351 & 0.625 & 0.326 & 0.616 \\ \hline 
Delhi & 0.340 & 0.622 & 0.323 & 0.619 \\ \hline 
Rajasthan & 0.332 & 0.618 & 0.282 & 0.599 \\ \hline 
Uttar Pradesh & 0.327 & 0.616 & 0.287 & 0.601 \\ \hline 
Bihar & 0.319 & 0.614 & 0.273 & 0.596 \\ \hline 
Sikkim & 0.323 & 0.620 & 0.251 & 0.591 \\ \hline 
Arunachal Pradesh & 0.324 & 0.616 & 0.294 & 0.604 \\ \hline 
Nagaland & 0.233 & 0.583 & 0.219 & 0.579 \\ \hline 
Manipur & 0.193 & 0.566 & 0.183 & 0.564 \\ \hline 
Mizoram & 0.269 & 0.597 & 0.243 & 0.588 \\ \hline 
Tripura & 0.295 & 0.607 & 0.268 & 0.596 \\ \hline 
Meghalaya & 0.264 & 0.594 & 0.221 & 0.579 \\ \hline 
Assam & 0.297 & 0.607 & 0.267 & 0.597 \\ \hline 

%

West Bengal & 0.369 & 0.635 & 0.338 & 0.622 \\ \hline 
Jharkhand & 0.344 & 0.624 & 0.299 & 0.607 \\ \hline 
Orissa & 0.355 & 0.627 & 0.323 & 0.615 \\ \hline 
Chattisgarh & 0.364 & 0.631 & 0.339 & 0.622 \\ \hline 
Madhya Pradesh & 0.363 & 0.630 & 0.326 & 0.616 \\ \hline 
Gujarat & 0.330 & 0.620 & 0.296 & 0.607 \\ \hline 
Daman \& Diu & 0.355 & 0.629 & 0.304 & 0.610 \\ \hline 
D \& N Haveli & 0.340 & 0.626 & 0.270 & 0.599 \\ \hline 
Maharashtra & 0.395 & 0.643 & 0.358 & 0.628 \\ \hline 
Andhra Pradesh & 0.373 & 0.635 & 0.342 & 0.623 \\ \hline 
Karnataka & 0.390 & 0.641 & 0.346 & 0.624 \\ \hline 
Goa & 0.317 & 0.611 & 0.300 & 0.605 \\ \hline 
Lakshadweep & 0.363 & 0.633 & 0.306 & 0.611 \\ \hline 
Kerala & 0.414 & 0.648 & 0.381 & 0.635 \\ \hline 
Tamil Nadu & 0.358 & 0.630 & 0.333 & 0.621 \\ \hline 
Pondicherry & 0.347 & 0.625 & 0.318 & 0.615 \\ \hline 
A \& N Island & 0.362 & 0.632 & 0.336 & 0.622 \\ \hline 
&&&& \\ \hline

\end{tabular}
\end{table}

\begin{table}
\caption{Estimated values of $g$ and $k$-indices for per capita consumption 
expenditure for India for 2004-2005 \&  2011-2012. Data is taken from 
Ref.~\cite{NSSO} and Ref.~\cite{chakrabarti2016quantifying}}
\label{tab:in_consu_g_k_2}

\begin{tabular}{|c|c|c|c|c|c|}
\hline

State &  \multicolumn{2}{c|}{2004-2005}   & 
\multicolumn{2}{c|}{2011-2012}  \\ \cline{2-5} 
& $g$ & $k$ &  $g$ & $k$  
\\ \hline 

Jammu \& Kashmir & 0.256 & 0.590 & 0.310 & 0.609 \\ \hline
Himachal Pradesh & 0.322 & 0.615 & 0.336 & 0.620 \\ \hline
Punjab & 0.318 & 0.614 & 0.334 & 0.620 \\ \hline
Chandigarh & 0.380 & 0.641 & 0.378 & 0.637 \\ \hline
Uttaranchal & 0.313 & 0.612 & 0.350 & 0.627 \\ \hline
Haryana & 0.310 & 0.611 & 0.365 & 0.631 \\ \hline
Delhi & 0.371 & 0.638 & 0.382 & 0.638 \\ \hline
Rajasthan & 0.310 & 0.609 & 0.332 & 0.619 \\ \hline
Uttar Pradesh & 0.320 & 0.614 & 0.357 & 0.628 \\ \hline
Bihar & 0.271 & 0.595 & 0.286 & 0.601 \\ \hline
Sikkim & 0.292 & 0.605 & 0.243 & 0.586 \\ \hline
Arunachal Pradesh & 0.541 & 0.690 & 0.371 & 0.637 \\ \hline
Nagaland & 0.222 & 0.579 & 0.241 & 0.587 \\ \hline
Manipur & 0.192 & 0.567 & 0.220 & 0.577 \\ \hline
Mizoram & 0.264 & 0.596 & 0.259 & 0.593 \\ \hline
Tripura & 0.312 & 0.610 & 0.290 & 0.606 \\ \hline
Meghalaya & 0.239 & 0.583 & 0.263 & 0.595 \\ \hline
Asaam & 0.288 & 0.601 & 0.309 & 0.610 \\ \hline

%

West Bengal & 0.347 & 0.625 & 0.387 & 0.643 \\ \hline
Jharkhand & 0.327 & 0.617 & 0.341 & 0.623 \\ \hline
Orissa & 0.352 & 0.624 & 0.347 & 0.625 \\ \hline
Chattisgarh & 0.346 & 0.624 & 0.367 & 0.632 \\ \hline
Madhya Pradesh & 0.337 & 0.620 & 0.366 & 0.632 \\ \hline
Gujarat & 0.347 & 0.626 & 0.345 & 0.624 \\ \hline
Daman \& Diu & 0.324 & 0.618 & 0.273 & 0.598 \\ \hline
D \& N Haveli & 0.362 & 0.636 & 0.335 & 0.622 \\ \hline
Maharashtra & 0.415 & 0.654 & 0.391 & 0.641 \\ \hline
Andhra Pradesh & 0.352 & 0.626 & 0.345 & 0.624 \\ \hline
Karnataka & 0.378 & 0.636 & 0.399 & 0.643 \\ \hline
Goa & 0.322 & 0.618 & 0.306 & 0.610 \\ \hline
Lakshadweep & 0.325 & 0.614 & 0.396 & 0.644 \\ \hline
Kerala & 0.385 & 0.639 & 0.431 & 0.655 \\ \hline
Tamilnadu & 0.374 & 0.638 & 0.357 & 0.628 \\ \hline
Pondicherry & 0.374 & 0.638 & 0.339 & 0.619 \\ \hline
A \& N Islands & 0.350 & 0.629 & 0.347 & 0.623 \\ \hline
India Total & 0.366 & 0.631 & 0.378 & 0.637 \\ \hline

\end{tabular}

\end{table}

\begin{table}
\caption{Estimated values of $g$ and $k$-indices for per capita consumption 
expenditure for Brazil for 2004-2005 \&  2008-2009. Data is taken from 
Ref.~\cite{chakrabarti2016quantifying} \& Ref.~\cite{Brazil0203,Brazil0809}.}
\label{tab:br_consu_g_k}

\begin{tabular}{|c|c|c|c|c|c|}
\hline

State &  \multicolumn{2}{c|}{2002-2003}   & 
\multicolumn{2}{c|}{2008-2009}  \\ \cline{2-5} 
& $g$ & $k$ &  $g$ & $k$  \\ \hline 

Rond\^{o}nia & 0.535 & 0.698 & 0.498 & 0.682 \\ \hline
Acre & 0.570 & 0.715 & 0.484 & 0.678 \\ \hline
Amazonas & 0.549 & 0.704 & 0.504 & 0.686 \\ \hline
Roraima & 0.529 & 0.696 & 0.558 & 0.706 \\ \hline
Par\'{a} & 0.509 & 0.687 & 0.538 & 0.698 \\ \hline
Amap\'{a} & 0.510 & 0.690 & 0.537 & 0.698 \\ \hline
Tocantins & 0.569 & 0.714 & 0.498 & 0.681 \\ \hline
Maranh\~{a}o & 0.502 & 0.686 & 0.524 & 0.692 \\ \hline
Piau\'{i} & 0.557 & 0.706 & 0.498 & 0.680 \\ \hline
Cear\'{a} & 0.571 & 0.711 & 0.514 & 0.686 \\ \hline
Rio Grande do Norte & 0.558 & 0.707 & 0.501 & 0.684 \\ \hline
Para\'{i}ba & 0.538 & 0.699 & 0.543 & 0.698 \\ \hline
Pernambuco & 0.558 & 0.706 & 0.532 & 0.695 \\ \hline
Alagoas & 0.583 & 0.719 & 0.541 & 0.698 \\ \hline
Sergipe & 0.518 & 0.692 & 0.512 & 0.687 \\ \hline

%
%

Bahia & 0.584 & 0.717 & 0.540 & 0.698 \\ \hline
Minas Gerais & 0.528 & 0.693 & 0.508 & 0.684 \\ \hline
Esp\'{i}rito Santo & 0.535 & 0.699 & 0.511 & 0.686 \\ \hline
Rio de Janeiro & 0.591 & 0.726 & 0.551 & 0.703 \\ \hline
S\~{a}o Paulo & 0.516 & 0.690 & 0.486 & 0.677 \\ \hline
Paran\'{a} & 0.519 & 0.691 & 0.470 & 0.671 \\ \hline
Santa Caterina & 0.465 & 0.668 & 0.498 & 0.680 \\ \hline
Rio Grande do Sul & 0.534 & 0.696 & 0.483 & 0.676 \\ \hline
Mato Grosso do Sul & 0.505 & 0.685 & 0.497 & 0.683 \\ \hline
Mato Grosso & 0.513 & 0.687 & 0.488 & 0.677 \\ \hline
Go\'{i}as & 0.506 & 0.685 & 0.523 & 0.689 \\ \hline
Distrito Federal & 0.590 & 0.725 & 0.564 & 0.715 \\ \hline
rural & 0.514 & 0.689 & 0.528 & 0.693   \\ \hline
urban & 0.568 & 0.711 & 0.533 & 0.695  \\ \hline
all Brazil & 0.478 & 0.679 & 0.507 & 0.683 \\ \hline

\end{tabular}
\end{table}

\begin{table}
\caption{Estimated values of $g$ and $k$-indices for per capita consumption 
expenditure for Italy for several years. Data is taken from 
Ref.~\cite{chakrabarti2016quantifying} \& Ref.~\cite{Italydata}.}
\label{tab:it_consu_g_k}

\begin{tabular}{|c|c|c|c|}
\hline

Year &   $g$ & $k$  \\ \hline 

1980 & 0.307 & 0.608 \\ \hline
1981 & 0.298 & 0.605 \\ \hline
1982 & 0.296 & 0.604 \\ \hline
1983 & 0.296 & 0.604 \\ \hline
1984 & 0.302 & 0.607 \\ \hline
1986 & 0.297 & 0.604 \\ \hline
1987 & 0.333 & 0.619 \\ \hline

\end{tabular}
\begin{tabular}{|c|c|c|c|}
\hline

Year &   $g$ & $k$  \\ \hline 

1989 & 0.288 & 0.602 \\ \hline
1991 & 0.285 & 0.601 \\ \hline
1993 & 0.297 & 0.606 \\ \hline
1995 & 0.305 & 0.608 \\ \hline
1998 & 0.316 & 0.612 \\ \hline
2000 & 0.308 & 0.610 \\ \hline
2002 & 0.317 & 0.613 \\ \hline

\end{tabular}
\begin{tabular}{|c|c|c|c|}
\hline

Year &   $g$ & $k$  \\ \hline 

2004 & 0.305 & 0.609 \\ \hline
2006 & 0.290 & 0.603 \\ \hline
2008 & 0.278 & 0.598 \\ \hline
2010 & 0.294 & 0.604 \\ \hline
2012 & 0.292 & 0.604 \\ \hline
&& \\ \hline
&& \\ \hline

\end{tabular}
\end{table}

\begin{table}[h]
\caption{Estimated values of $g$ and $k$-indices from  income distribution for 
USA for several years. Data is taken from Ref.~\cite{IRS}.}
\label{tab:us_income}

\begin{tabular}{|c|c|c|c|}
\hline

Year &  $g$ & $k$  \\ \hline 

1996 & 0.5519 & 0.6994 \\ \hline
1997 & 0.5600 & 0.7015 \\ \hline
1987 & 0.5848 & 0.7071 \\ \hline
1999 & 0.5760 & 0.7078 \\ \hline
2000 & 0.5842 & 0.7114 \\ \hline
2001 & 0.5612 & 0.7029 \\ \hline
2002 & 0.5501 & 0.6990 \\ \hline
2003 & 0.5551 & 0.7016 \\ \hline

\end{tabular}
\begin{tabular}{|c|c|c|c|}
\hline

Year &  $g$ & $k$  \\ \hline

2004 & 0.5398 & 0.6926 \\ \hline
2005 & 0.6042 & 0.7161 \\ \hline
2006 & 0.5945 & 0.7129 \\ \hline
2007 & 0.6008 & 0.7154 \\ \hline
2008 & 0.5840 & 0.7095 \\ \hline
2009 & 0.5330 & 0.6918 \\ \hline
2010 & 0.5735 & 0.7066 \\ \hline
2011 & 0.5780 & 0.7086 \\ \hline 

\end{tabular}
\end{table}

\begin{table}[h]
\caption{Estimated values of  $g$ and $k$-indices from  vote distribution for  
several countries with \textit{First-past-the-post} election system for several 
years. Data sources: India~\cite{ECI}, UK~\cite{UKElec}, Canada~\cite{CAElec}, 
Bangladesh~\cite{BDElec}, Tanzania~\cite{TZElec}.}
\label{tab:fptp_vote}

\begin{tabular}{|r|c|c|}
\hline

Country/State \& Year &   $g$ & $k$  \\ \hline 
India 1980 & 0.726 & 0.791 \\ \hline
1984 & 0.786 & 0.830 \\ \hline
1989 & 0.805 & 0.839 \\ \hline
1991 & 0.844 & 0.864 \\ \hline
1996 & 0.882 & 0.895 \\ \hline
1998 & 0.749 & 0.801 \\ \hline
1999 & 0.732 & 0.792 \\ \hline
2004 & 0.752 & 0.810 \\ \hline
2009 & 0.804 & 0.846 \\ \hline \hline 

West Bengal 1972 & 0.396 & 0.633 \\ \hline
1977 & 0.529 & 0.689 \\ \hline
1982 & 0.526 & 0.683 \\ \hline
1987 & 0.617 & 0.726 \\ \hline
1991 & 0.654 & 0.750 \\ \hline
1996 & 0.692 & 0.772 \\ \hline
2001 & 0.623 & 0.738 \\ \hline
2011 & 0.640 & 0.747 \\ \hline \hline 
%
%

Uttar Pradesh 1996 & 0.722 & 0.785 \\ \hline
2002 & 0.723 & 0.793 \\ \hline
2012 & 0.762 & 0.818 \\ \hline \hline 

Madhya Pradesh 1990 & 0.792 & 0.837 \\ \hline
1998 & 0.693 & 0.773 \\ \hline
2008 & 0.752 & 0.818 \\ \hline \hline 

Andhra Pradesh 1999 & 0.707 & 0.776 \\ \hline
2004 & 0.676 & 0.761 \\ \hline
2009 & 0.755 & 0.815 \\ \hline \hline

Bihar 1990 & 0.818 & 0.843 \\ \hline
1995 & 0.821 & 0.847 \\ \hline
2010 & 0.705 & 0.795 \\ \hline
%

\end{tabular}
\begin{tabular}{|r|c|c|}
\hline
Country \& Year &   $g$ & $k$  \\ \hline 

UK 1970 & 0.334 & 0.615 \\ \hline
1979 & 0.468 & 0.669 \\ \hline
1983 & 0.411 & 0.643 \\ \hline
1987 & 0.378 & 0.636 \\ \hline
1992 & 0.512 & 0.686 \\ \hline
1997 & 0.588 & 0.721 \\ \hline
2001 & 0.531 & 0.698 \\ \hline
2005 & 0.535 & 0.698 \\ \hline
2010 & 0.580 & 0.718 \\ \hline \hline

Canada 2000 & 0.597 & 0.728 \\ \hline
2004 & 0.558 & 0.707 \\ \hline
2006 & 0.530 & 0.695 \\ \hline
2008 & 0.517 & 0.693 \\ \hline
2011 & 0.537 & 0.696 \\ \hline \hline 

Bangladesh 1973 & 0.592 & 0.731 \\ \hline
2000 & 0.687 & 0.761 \\ \hline \hline 

Tanzania 2005 & 0.721 & 0.787 \\ \hline
2010 & 0.653 & 0.746 \\ \hline 
& & \\ [30ex] \hline

\end{tabular}
\end{table}

\begin{table}[h]
\caption{Estimated values of  $g$ and $k$-indices from  vote distribution for  
several countries with \textit{Open list proportional} election system for 
several years. Data is taken from Ref.~\cite{chatterjee2013universality}.}
\label{tab:op_vote}

\begin{tabular}{|l|c|c|c|}
\hline

Country & Year &   $g$ & $k$  \\ \hline 
Italy & 1976 & 0.5593 & 0.7077 \\\hline
Italy & 1979 & 0.5463 & 0.7014 \\\hline
Italy & 1987 & 0.5720 & 0.7144 \\\hline \hline 
Netherlands & 2010 & 0.9406 & 0.9214 \\\hline
Netherlands & 2012 & 0.9250 & 0.9071 \\\hline \hline 
Sweden & 2006 & 0.6903 & 0.7650 \\\hline
Sweden & 2010 & 0.7374 & 0.7842 \\\hline

\end{tabular}
\end{table}

\begin{table}
\caption{Estimated values of $g$ and $k$-indices from  population distribution 
for cities/municipalities several countries. Data is taken from 
Ref.~\cite{ghosh2014zipf}.}
\label{tab:city}

\begin{tabular}{|l|c|c|c|}
\hline

Country & Year &   $g$ & $k$  \\ \hline 
Brazil & 2012 & 0.7270 & 0.7795 \\\hline
Spain & 2011 & 0.8661 & 0.8560 \\\hline
Japan & 2010 & 0.7192 & 0.7738 \\\hline
\end{tabular}
\end{table}

\end{document}